\documentclass[twocolumn,superscriptaddress,showpacs,aps,prb]{revtex4}
\usepackage{makeidx}
\usepackage{bm}
\usepackage{graphics}
\usepackage[dvips]{epsfig}
\usepackage{mathtext}
\usepackage{amsmath}
\newcounter{fig}

\begin{document}
\title{Optical far-infrared properties of   graphene monolayer
and  multilayers }
\author{L.A. Falkovsky} \affiliation{L.D. Landau
Institute for Theoretical Physics, Moscow 117334, Russia}
\affiliation{Institute of the High Pressure Physics, Troitsk
142190, Russia}
\author{C.C. Persheguba }
\affiliation{ Moscow Institute of Physics and Technology,
Dolgoprudniy 147100, Russia}
\date{\today}
\begin{abstract}
We analyze the  features   of  the graphene mono- and multilayer
reflectance in the far-infrared region as  a function of
frequency, temperature, and carrier density taking  the intraband
conductance and the interband electron absorbtion into account.
 The dispersion of plasmon mode of
the  multilayers  is calculated  using Maxwell's equations with
the influence of retardation included. At low temperatures and
high electron densities, the reflectance of multilayers as a
function of frequency has the sharp downfall and the subsequent
deep well due to the threshold of electron interband absorbtion.
\end{abstract}

\maketitle
%\section{Introduction}
Monolayer and bilayer graphenes \cite{Novo,ZSA,NF} are gapless
two-dimensional (2D) semiconductors \cite{DD,An,MF} whereas its 3D
predecessor---graphite is a semimetal \cite{W,SW,McCl}. Hence the
dimensionality effects for the unique substance can be studied
\cite{Fe}. Monolayer graphene has a very simple electron band
structure. Near the  energy $\varepsilon=0$, the energy bands are
cones $\varepsilon_{1,2}(\mathbf{p})=\pm vp$ at the $ K$ points in
the 2D Brillouin zone with the constant velocity parameter
$v=10^8$ cm/s.  Such a degeneration is conditioned by symmetry
because the small group $C_{3v}$  of the $K$ points has
two-dimensional representation.

While the carrier concentration is decreasing in the field gate
 experiment, the graphene conductivity at low temperatures
 goes to the finite minimal values \cite{Novo,ZSA}.
Much theoretical efforts \cite{Li,LFS,TTT,Zk} have been devoted to
evaluate the minimal conductivity in different approaches.
Theoretical \cite{An1,NMD,PGC} and experimental researches  show
that the main mechanism of the carrier relaxation is provided
 by the charged impurities and gives the collision  rate
  $\tau^{-1}\sim 2\pi^2e^4n_{imp}/\hbar\epsilon_g^2\varepsilon$, where
 $\epsilon_g$ is the dielectric constant of graphene, $\varepsilon$ is
 the characteristic electron energy (of the order of  the Fermi energy
 or temperature),
 and $n_{imp}$ is the density of charged impurities per the unit surface.
   Plasmons in graphene are considered in Refs. \cite{FV, HDS}.
   The optical visibility of both monolayer and bilayer graphene is
   studied in Ref. \cite{ARF} focusing on the  role of the
    underlying substrate.

In the present paper, we analyze  the spectroscopy of the graphene
monolayer and multilayers in the infra-red region. In order to
calculate the reflection coefficient for the multilayers, we
follow  the method  used in Ref. \cite{FM} and determine the
spectrum of electromagnetic excitations---plasmons. We use the
appropriate boundary conditions at  interfaces and the complex
conductivity $\sigma$ as a function of frequency $\omega$,
temperature $T$, and chemical potential $\mu$. The chemical
potential of ideal pure graphene equals to zero at any
temperature. With the help of the gate voltage, one can control
the density and type ($n$ or $p$) of carriers varying their
chemical potential.

The general expression for the conductivity used here is obtained
in our previous paper \cite{FV} and is valid under a restriction
that the collision rate of carriers is less than the frequency and
spatial dispersion of the electric ac field, $\tau^{-1}\ll\omega,
kv $.
 In   limiting cases,
 our result coincides  with the formulas  of Ref. \cite{GSC,Cs}.
%\section{Optical conductivity of graphene}
For high frequencies, when one can also ignore the spacial
dispersion of the ac field, $\omega\gg kv,\tau^{-1}$, the complex
conductivity [see  Eq. (8) in  Ref. \cite{FV}] is given by
\begin{equation}
  \sigma(\omega) = \frac{e^2\omega}{i\pi\hbar}\left
  [\int\limits_{-\infty}^{+\infty} d\varepsilon\frac{|\varepsilon |}{\omega ^2}
   \frac{df_0 (\varepsilon)}{d\varepsilon}- \int\limits_0^{+\infty}
  d\varepsilon\frac{f_0 (-\varepsilon)-f_0(\varepsilon)}{(\omega+i\delta)^2 -
  4\varepsilon^2}\right]\, .
  \label{sigma}
\end{equation}
 Here, the first term
corresponds to the intraband electron--photon scattering
processes. One can  obtain it from  the Drude--Boltzmann
expression (for a case $1/\tau=0$) and write explicitly:
\begin{equation}
     \sigma ^{intra}(\omega) =i\frac{2e^2T}
     {\pi\hbar\omega}
\ln{[2\cosh(\mu /T)]}
 \label{sigm}    \, .
 \end{equation}
The second term in Eq. (1), where $\delta\to 0$ is the
infinitesimal quantity determining the bypass around the integrand
pole, owes its origin to the direct interband electron
transitions. The real part of this contribution is reduced to the
expression for the absorbed energy due to the  interband
transitions.
 Since there is no
 gap between the conduction band and valence band, these two terms
 can compete  and the interband contribution becomes  larger at
 high frequencies $\omega>T, \mu$. In the opposite case,
 the intraband contribution plays the leading role.

 The difference of the Fermi functions in the second integrand  equals to
 $$G(\varepsilon)=\frac{\sinh(\varepsilon/T)}{\cosh(\mu/T)+\cosh(\varepsilon/T)}\, .$$
Extracting the principal value of the integral, we arrive at the
integral without singularities and write the interband
conductivity in the form available for numerical calculations:
  \begin{equation}
    \sigma^{inter}(\omega) =
    \frac{e^2}{4\hbar}\left[G(\omega/2)
     -\frac{4\omega}{i\pi}\int\limits_0^{+\infty}
  d\varepsilon\frac{G(\varepsilon)-G(\omega/2)}{\omega^2 -
  4\varepsilon^2}\right]\, .
 \label{equat6} \end{equation}
Here, the first term  is given asymptotically by
\begin{eqnarray}
G(\omega/2)= \left\{
\begin{array}{ll}
\displaystyle\tanh(\omega/4T) , & \mu\ll T\, , \\
 \displaystyle\theta(\omega-2\mu), & \mu\gg T\, ,
\end{array}
\right. \label{g} \end{eqnarray} where the step-function
$\theta(\omega-2\mu)$ expresses the condition for  the interband
electron transitions at low temperatures. The integral in Eq.
(\ref{equat6}) represents the imaginary interband correction to
the intraband conductivity.

\begin{figure}[h]
\noindent\centering{
\includegraphics[width=90mm]{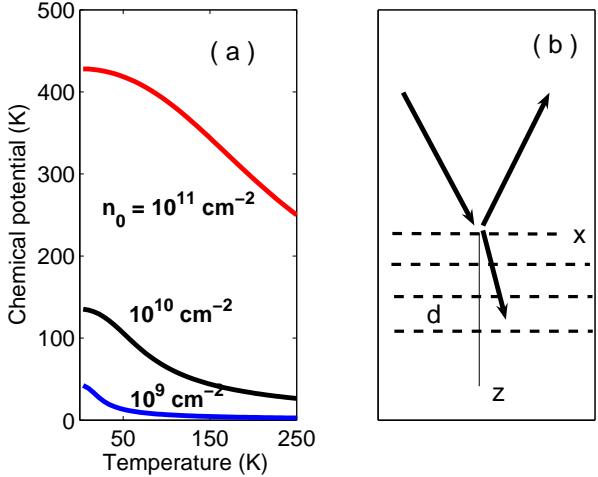}
} \caption{(a) Chemical potential (in K) as a function of
temperature at carrier densities noted at curves. (b) Multilayers
sample and geometry of wave scattering.} \label{mu}
\end{figure}

  By using the gate
voltage, one can control the density of electrons ($n_0$) or holes
($-n_0$). Then the chemical potential is determined by the
condition
\begin{eqnarray}
n_0 =\frac{2}{\pi(\hbar
v)^2}\int\limits_0^{+\infty}\varepsilon[f_0(\varepsilon-\mu)-
f_0(\varepsilon+\mu)] d\varepsilon\, . \label{den}\end{eqnarray}
From this expression and Fig. \ref{mu}(a), one can see that the
chemical potential
  goes to zero while the temperature increases.

%\section {Spectroscopy of graphene layers}

In order to calculate the graphene reflectance, we apply Maxwell's
equations
\begin{equation}
\label{maxwell} \nabla (\nabla \cdot {\bf E}) -\nabla^2 {\bf
E}=\epsilon_0\frac{\omega^2}{c^2}{\bf E}+ \frac{4\pi
i\omega}{c^2}{\bf j}\, ,
\end{equation}
where $\epsilon_0$ is the ion contribution into the dielectric
constant and ${\bf j}$ is the conductivity current. We consider
the case of the $p$-polarization, when the field ${\bf E}$ lies in
the $xz$ plane and the current ${\bf j}$ has  only the in-layer
$x$ component (see Fig \ref{mu}b).

{\it (i) Optics of a monolayer.}
 Consider the graphene monolayer at $z=0$ with
 $\epsilon_0=\epsilon _g$
 deposited on the substrate ($z>0$) with the dielectric constant
$\epsilon_0=\epsilon _s$. In the vacuum, $z<0$, the ac field is
given by the sum of incident and reflected waves and by the
transmitted wave in the substrate. In the geometry considered, the
current in graphene monolayer can be written in the form
\begin{equation} j_x=\sigma
\delta(z)E_x\, .\label{car} \end{equation}

Making use of the Fourier transformations with respect to the $x$
coordinate, ${\bf E}\propto e^{i k_xx}$,  we rewrite the Maxwell
equations (\ref{maxwell}) as follows
\begin{eqnarray}\label{mxeqn1}
\begin{array}{cc}
\displaystyle
ik_x\frac{dE_z}{dz}-\frac{d^2E_x}{dz^2}-\epsilon_0\frac{\omega^2}{c^2}E_x
=\frac{4\pi i\omega}{c^2}j_x \, ,&\\
\displaystyle
ik_x\frac{dE_x}{dz}+(k_x^2-\epsilon_0\frac{\omega^2}{c^2})E_z=0 \,
.&
\end{array}
\end{eqnarray}
\begin{figure}[h]
\noindent\centering{
\includegraphics[width=90mm]{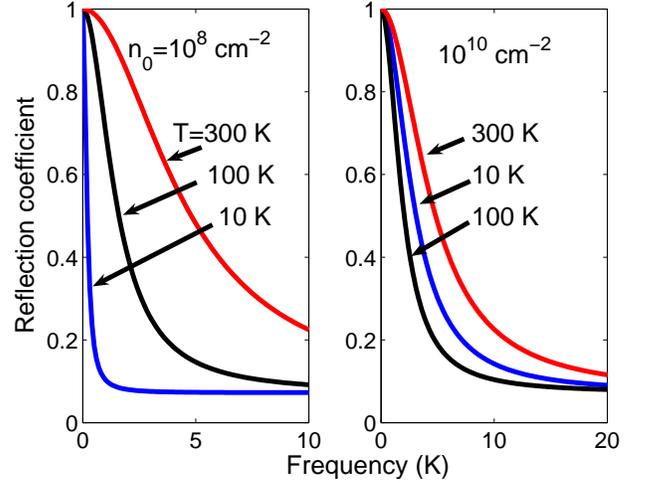}
} \caption{Reflectance from the graphene monolayer with  carrier
densities $n_0=10^{8}$ cm$^{-2}$ (left) and $n_0=10^{10}$
cm$^{-2}$ (right) versus the frequency  at temperatures noted at
the curves; normal incidence.} \label{figfr}
\end{figure}

The boundary conditions for these equations  at  $z=0$
   are  the continuity of the field component $E_x$ and the
jump of the electric-induction $z$ component  $\epsilon E_z$ at
sides of the monolayer:
 \begin{eqnarray}\label{jump}
\epsilon_sE_z|_{z=+0}- E_z|_{z=-0}=4\pi
\int_{-0}^{+0}\rho(\omega,k_x,z)dz\, .
\end{eqnarray}
The carrier density is connected to the current in Eq. (\ref{car})
according to the continuity equation
$$\rho(\omega,k_x,z)=j_x(\omega,k_x,z)k_x/\omega.$$
Substituting $E_z$ from the second Eq. $(\ref{mxeqn1})$ into
(\ref{jump}), we find the second boundary condition
\begin{equation}\frac{\epsilon_s}{k_s^2}\frac{dE_x}{dz}|_{z=+0}-
\frac{1}{(k_z^{i})^2}\frac{dE_x}{dz}|_{z=-0}=\frac{4\pi
\sigma(\omega)}{i\omega}E_x|_{z=0}\, , \label{bc}\end{equation}
 where
$$k_s=\sqrt{\epsilon_s(\omega/c)^2-k_x^2},\quad
k_z^i=\sqrt{(\omega/c)^2-k_x^2}\, .$$

Using the boundary conditions, we find the reflection amplitude
 \begin{equation}
 r =  \frac{1-C}{1+C}\, ,
\label{tr}\end{equation} where
$C=k_z^i[4\pi\sigma(\omega)/\omega+(\epsilon_s/k_s)]\, .$

\begin{figure}[h]
\noindent\centering{
\includegraphics[width=90mm]{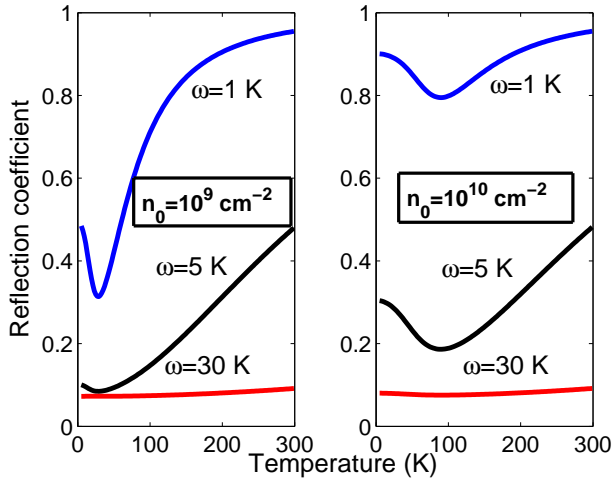}
} \caption{Reflectance from the graphene monolayer
 with  carrier concentrations $n_0=10^{9}$ cm$^{-2}$ (left) and
$10^{10}$ cm$^{-2}$ (right) versus temperature for frequencies
noted at  curves; normal incidence.} \label{figtem}
\end{figure}

The reflection coefficient calculated with the help of Eqs.
(\ref{sigma})--(\ref{den})  and (\ref{tr}) for  normal incidence
is shown in Figs. \ref{figfr} and \ref{figtem} as a function of
frequency, temperature, and carrier concentration. Notice the
different temperature behavior of reflectance from the samples
containing the low and high carrier densities. Reflectance for the
lower carrier density (Fig. \ref{figfr}, left) is larger at higher
temperatures. This corresponds to the increase in the intraband
conductivity (giving the main contribution here) with temperature
[see Eq. (\ref{sigm}) for $\mu<< T$].  At the larger carrier
concentration (Fig. \ref{figfr}, right), the chemical potential
(decreasing with temperature, see Fig. \ref{mu}a) plays the
important role appearing in Eq. (\ref{sigm}) instead of $T$.
Therefore, the curve for lowest temperature $T=10$ K is between
the curves for $T=100$ K and $300 $ K in this case. The
temperature dependencies of reflectance are not monotonic as
clearly seen in Fig. \ref{figtem}.

 The optical properties of the
graphene bilayer can be considered in a similar way. Here we do
not present the corresponding results and investigate another 3D
example---the graphene multilayers.

{\it (ii) Spectroscopy of graphene multilayers.} Let the
multilayers cross the $z$ axis at points $z_n=nd$, where $d$ is
the distance between the layers (see Fig. \ref{mu}b). Such a
system can be considered as a model of graphite since the distance
$d=3 \AA$ in graphite is larger than the interatomic distance in
the layer. So we describe the carrier interaction in the presence
of  ac electric field with the help of self-consistent Maxwell's
equations (\ref{maxwell}). For the $x$ component of the field
$E_x$, they give
\begin{equation}\label{max1}
 \left(\frac{d^2}{dz^2}+k_s^2+2k_s {\cal
D}\sum_{n}\delta(z-nd)\right) E_x=0\, ,
\end{equation}
where $ {\cal D}=2i\pi\sigma(\omega)k_s/\epsilon_g\omega\, . $

For the infinite number of layers in the stack, the solutions of
Eq. (\ref{max1}) represent two Bloch states
\begin{eqnarray}\nonumber
e_{1,2}(z)=e^{\pm ik_znd}\{\sin{k_s(z-nd)}- e^{\mp ik_zd}\times\\
\sin{k_s[z-(n+1)d]}\},\quad nd<z<(n+1)d \label{tsol}\end{eqnarray}
with  the quasi-momentum $k_z$ determined from
 the dispersion equation
\begin{equation}\label{dr}
\cos{k_zd}=\cos{k_s d}-{\cal D}\sin{k_s d}\, .
\end{equation}
The dispersion equation describes the electric field excitations
of the system, i.e., plasmons. The quasi-momentum $k_z$ can be
restricted to the Brillouin half-zone $0<k_z<\pi/d$, if the
parameter ${\cal D}$ is real. In the general case, while taking
 the interband absorbtion into account,
 we fix the choice of the eigen-functions in
Eq.~(\ref{tsol}) by the condition ${\rm Im}\, {k_z}>0$ so that the
solution $e_{1}$ decreases in the positive direction $z$.
\begin{figure}[]
\noindent\centering{
\includegraphics[width=90mm]{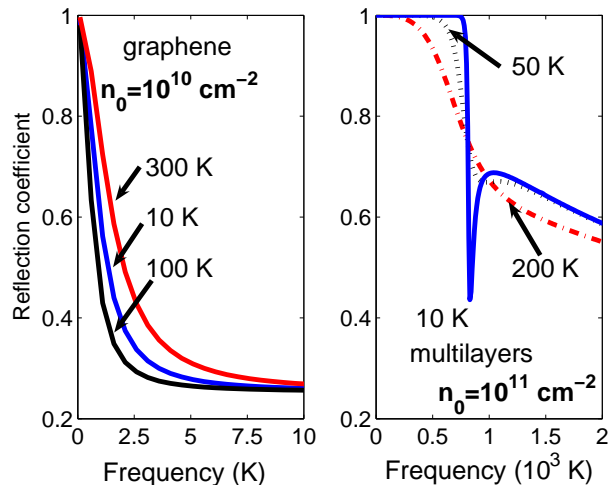}
} \caption{Reflectance from the graphene monolayer with the
carrier density  $n_0=10^{10}$ cm$^{-2}$ (left) and  multilayers
 with $n_0=10^{11}$ cm$^{-2}$ in a layer and
distance $d=3 \AA$ between layers (right); temperatures are noted
at curves, the incidence angle is 80$^{\rm 0}$.} \label{ref3-2d}
\end{figure}

Reflectance from the multilayers occupied the semi-space
 $z>0$ can be calculated similarly to the reflectance of a monolayer.
 The electric field is given by the decreasing solution
  $e_1$ inside the sample and by the sum of   incident and
  reflected waves
  in the vacuum, $z<0$, with the same value of the component $k_x$.

Using the boundary condition, Eq. (\ref{bc}), with the dielectric
constant  of graphene $\varepsilon_g$ instead of $\varepsilon_s$,
we find the reflected amplitude
$$r =\frac{i\sin(k_s d)-Z}{i\sin(k_s d)+Z}\, ,$$
where
$$Z=\epsilon_g\frac{k_z^i}{k_s}\left[\cos(k_s d)-e^{-ik_zd}\right]\, ,$$
 $k_z^i$ is the normal component of
  wavevector in the vacuum
  and $k_z$ is the quasi-momentum determined by the
  dispersion equation(\ref{dr})
 at  fixed values of $\omega$ and $k_x$.

 In Fig. \ref{ref3-2d},  the reflection coefficient calculated for
 multilayers
is shown  in comparison with the reflection coefficient of the
monolayer. The left panel in this figure differs from the right
one in Fig. \ref{figfr} only in the incidence angle which is now
taken to be  80$\,^{\rm o}$ in order to emphasize the multilayer
features.  The main of them is the sharp downfall of reflectance
at low temperatures (see the right panel in Fig. \ref{ref3-2d}).
This is the threshold effect of the direct interband transitions
at $\omega\ge 2\mu$, which is sharp when temperature $T\to 0$ [see
Eq. (\ref{g})]. Just after  the downfall, the reflectance has  the
deep well which disappears while the temperature increases. Then
the downfall becomes more smooth. Notice that observations of the
absorbtion threshold provide a direct method of carrier density
 characterization  of graphene $n_0=(\mu/\hbar v)^2/\pi$.

In conclusion, we have developed the detailed microscopic theory
of   the graphene mono- and multilayer spectroscopy. We have shown
that the nonmonotonic temperature behavior of reflectance from the
monolayer in the infra-red region is expected.  We have argued
that at low temperatures and high electron densities,
 the reflectance from multilayers  has the sharp
downfall with the subsequent deep well. They are caused by the
direct interband electron transitions.

This work is supported by the Russian Foundation for Basic
Research (grant No.07-02-00571).

\end{document}